\documentclass[11pt]{JHEP}
\usepackage{latexsym}
\usepackage{youngtab}
\usepackage{amssymb,amsfonts}
\usepackage{amsmath,amsthm,epsfig,euscript,array,cancel}
\font\mybb=msbm10 at 11pt
\def\bb#1{\hbox{\mybb#1}}

\preprint

\preprint{Dec. 7, 2015 }

\title{On section conditions of $E_{7(+7)}$ exceptional field theory and superparticle in ${\cal N}=8$ central charge superspace}

\author{
Igor Bandos
\\ \vspace{0.8cm}
Department of
Theoretical Physics, University of the Basque Country UPV/EHU,
\\ P.O. Box 644, 48080 Bilbao, Spain
 \\ and
IKERBASQUE, Basque Foundation for Science, 48011, Bilbao, Spain
}

\date{December  2014--  December 2015}

\abstract{We study the properties of section conditions of the $E_{7(+7)}$ exceptional field theory from the perspective of superparticle model in  ${\cal N}=8$ $D=4$ superspace enlarged by additional bosonic coordinates related to the central charge of the maximal supersymmetry superalgebra. In particular, the superparticle model suggests that only a part of the section conditions corresponding to generators of $SU(8)$ subgroup of $E_{7(+7)}$ is independent, and we show that this part  indeed suffices to obtain the (classical counterpart of the) general solution of the strong section condition. }


\begin{document}

\maketitle

\setcounter{page}2

\section{Introduction}

The present stage of development of String/M-theory is characterized, at least partially, by exploiting the possibility to make manifest duality symmetries by introducing additional bosonic coordinates. Double field  theory (DFT) \cite{Siegel:1993xq,Hull:2004in,Hull:2009mi,Coimbra:2011nw,Jeon:2011vx,Hohm:2013bwa,Hatsuda:2014qqa,Hatsuda:2015cia}  designed to have a manifest T-duality symmetry, characteristic for string theory (see e.g. \cite{Green:1987sp}), is formulated in the space with doubled number of (D=10) spacetime coordinates.
To make manifest the U-duality symmetry of M-theory \cite{Hull:1994ys} one develops
 ''$E_{n(n)}$ exceptional field theories'' (EFTs)
\cite{Hull:2007zu,Berman:2012vc,Coimbra:2012af,Hohm:2013pua,Hohm:2013vpa,Hohm:2013uia,Hohm:2014fxa,Godazgar:2014nqa,Musaev:2014lna,Ciceri:2014wya,Abzalov:2015ega,Musaev:2015ces} which are also formulated in spacetime with  additional bosonic coordinates $y^\Sigma$. In both cases the dependence of the fields on additional coordinates is restricted by the so--called {\it section conditions}
the strong version of which is imposed on a pair of any two functions of the theory.

A progress in comprehension of structure and clarification of the origin  of these mysterious conditions are one of the main goals of this paper.

In the case of double field theory the strong  section conditions are imposed on any pair of two functions of 2D coordinates
and are solved by the conditions that all the physical fields depend only on D of 2D bosonic coordinates. The manifest T-duality is provided by the freedom in choosing the set of these $D$ of the complete set of $2D$ coordinates, which is often referred to as  choice of the section (hence the name of section conditions).
An interesting approach relating section conditions of DFT to a very special gauge symmetry 
(`gauge coordinate transformations') was developed in \cite{Park:2013mpa,Lee:2013hma}. 

In 'exceptional field theory' the situation is a bit more complicated, but also more illustrative.
To make manifest $E_{n(n)}$ duality symmetry the number of spacetime coordinates is generically more
then doubled  and the origin of the corresponding section conditions is more obscure.

It this paper we try to understand better  the structure and meaning of section conditions by approaching them from the perspective of superparticle model in  central charge superspace.
We restrict ourselves by a particular but important and illustrative case of $E_{7(+7)}$ effective field theory which is formulated in the
space with $60=4+56$ bosonic coordinates $(x^\mu, y^\Sigma)$ and the first observation beyond our approach is that the additional 56 coordinates $y^\Sigma$ can be associated with central charges $Z_{\Sigma}=(\bar{Z}_{ij},{Z}^{ij})$ of the maximal $D=4$ ${\cal N}=8$ supersymmetry algebra\begin{eqnarray}\label{SUSYal}
\{ Q_{\alpha}^i, Q_{\beta}^j\}= \epsilon_{\alpha\beta}Z^{ij}
\; , \qquad \{ Q_{\alpha}^i, \bar{Q}_{\dot{\beta}j}\}= \delta_j^i\sigma^a_{\alpha\dot{\beta}}P_a
\; , \qquad \{  \bar{Q}_{\dot{\alpha}i},  \bar{Q}_{\dot{\beta}j}\}= \epsilon_{\dot\alpha\dot\beta}\bar{Z}_{ij}
\; . \qquad
\end{eqnarray}
The supergroup manifold $\Sigma_0^{(60|32)}$ corresponding to this superalgebra is called {\it central charge superspace}. The set of their coordinates contains, besides the coordinates of standards ${\cal N}=8$ $d=4$ superspace, also 56 bosonic coordinates $y^\Sigma=({y}^{ij},\bar{y}_{ij})$, which can be called {\it central charge coordinates}. The curved version of central charge superspace, $\Sigma^{(60|32)}$, was used for alternative  superspace formulation of $D=4$ ${\cal N}=8$ supergravity in \cite{Howe:1980th} (see also \cite{Howe:2015hpa}).

The use of (curved) central charge superspace allowed the authors of \cite{Howe:1980th}
to replace the superform generalization of 28  gauge fields of the supergravity multiplets and their 28 magnetic duals (which both appear in supergravity formulation in the standard ${\cal N}=8$ superspace \cite{Brink:1979nt,Howe:1981gz}) by the bosonic supervielbein forms corresponding to new coordinates,  $y^{\Sigma}=(y^{ij}, \bar{y}_{ij})$. To remove the unwanted extra degrees of freedom,  all the (super)fields are assumed in \cite{Howe:1980th,Howe:2015hpa} to be independent on  $y^{\Sigma}$, {\it i.e.} the conditions
\begin{eqnarray}\label{dy=0}
 \partial_{{\Sigma}} (\ldots)=0\;
\end{eqnarray}
are imposed\footnote{To be precise, \cite{Howe:2015hpa} uses the condition ${\cal D}_\Sigma (...)=0$ with  $SL(2,{\bb C})\times SU(8)$ covariant derivative ${\cal D}$.  }. With this condition  the model in central charge superspace is equivalent to  $D=4$ ${\cal N}=8$ supergravity \cite{Cremmer:1979up}.

From the above perspective one can state that
the $E_{7(+7)}$ EFT of \cite{Hohm:2013pua,Hohm:2013uia,Godazgar:2014nqa}  was formulated in a bosonic body of the central charge superspace. In it one  imposes, instead of (\ref{dy=0}),  the following {\it strong section conditions}
\begin{eqnarray}\label{sec=tdd=0}
& t_{E}{}^{{\Sigma}{\Pi}} \partial_{{\Sigma}} \otimes \partial_{{\Pi}}=0 & \; , \qquad \\
\label{sec=Sp=0}
& \Xi^{{\Sigma}{\Pi}} \partial_{{\Sigma}} \otimes \partial_{{\Pi}}=0  & . \qquad
\end{eqnarray}
We refer on the main text for the notation and more details (see sec. 2, Eqs. (\ref{sec=tdFdF2=0}), (\ref{sec=Sp2=0}) and below) and just notice
here that the 56$\times$ 56 matrix $\Xi$  (symplectic `metric') is antisymmetric   $\Xi^{{\Sigma}{\Pi}} =- \Xi^{{\Pi}{\Sigma}}$ while  $133$ matrices $t_{E}{}^{{\Sigma}{\Pi}}$ (related to the generators $t_{E\Lambda}{}^{\Pi}$ of $E_{7(+7)}$ by $t_{E}{}^{{\Sigma}{\Pi}} = \Xi^{{\Sigma}\Lambda} t_{E\Lambda}{}^{\Pi} $) are symmetric,
 $t_{E}{}^{{\Sigma}{\Pi}}=t_{E}{}^{{\Pi}{\Sigma}}$,
  and
the direct product symbol reflects the fact that these conditions are applied to any pair of the functions of the EFT. In addition, the weak section conditions
\begin{eqnarray}\label{sec=tdd=w}
t_{E}{}^{{\Sigma}{\Pi}} \partial_{{\Sigma}}  \partial_{{\Pi}}(\ldots)=0. \qquad
\end{eqnarray}
are obeyed by any function of EFT. Notice that this is a counterpart of the strong condition (\ref{sec=tdd=0}) while
the weak  counterpart of (\ref{sec=Sp2=0}) is satisfied identically due to antisymmetry of symplectic `metric', $  \Xi^{{\Sigma}{\Pi}} \partial_{{\Sigma}}\partial_{{\Pi}}\equiv 0$.

The general expectation is that the result of imposing the strong section condition is that all the fields depend on a smaller number $\tilde{n}$ of 'internal' coordinates $\tilde{y}^r$, schematically
\begin{eqnarray}\label{dy=Kdr}
 \partial_{{\Sigma}} (\ldots)= K_{\Sigma}{}^r \partial_r(\ldots)\; , \qquad \partial_r= \frac {\partial } {\partial \tilde{y}^r} \; , \qquad r=1,..., \tilde{n}\;
\end{eqnarray}
with some 56$\times \tilde{n}$ matrix $ K_{\Sigma}{}^r $. A possible choice of this latter defines a {\it section} (i.e. a particular solution of the section conditions).

In  the case of $E_{7(7)}$ EFT \cite{Hohm:2013pua,Hohm:2013uia,Godazgar:2014nqa} the solutions with   $\tilde{n}=n=7$ (we will discuss this below) and with  $\tilde{n}=6 $ are found and shown to correspond to the embedding of D=11 and D=10 type IIB supergravity into the EFT.
The general expectation is that  $\tilde{n}\leq n$ for any solution of the strong section conditions:  on physical/M-theoretic grounds one should not expect to have a solution allowing physical fields depend on more than 11 coordinates. However, to our best knowledge, this fact has not been proved till now
 \footnote{ The author thanks Henning Samtleben for a discussion on this and related issues.   }.
Below we will show that this is indeed the case. In achieving this the suggestions obtained by studying   superparticle in central charge superspace happens to be useful.

When particle mechanics in central charge superspace is considered, the momenta $p_{\Sigma}= (\bar{p}_{ij},p^{ij})$ conjugate to the central charge coordinates $y^{\Sigma}$ serves as prototypes   of the derivatives $\partial_\Sigma$. As $p_{\Sigma}$ are commutative canonical variables,  both the strong and the weak section conditions are represented in superparticle model by
\begin{eqnarray}\label{sec=tpp}
t_{E}{}^{{\Sigma}{\Pi}} \, p_{{\Sigma}}  \, p_{{\Pi}}=0. \qquad
\end{eqnarray}
Again, as in the case of weak section conditions, the 'classical' counterpart of the conditions
(\ref{sec=Sp2=0}) is satisfied identically due to antisymmetry of symplectic metric, $  \Xi^{{\Sigma}{\Pi}} p_{{\Sigma}}p_{{\Pi}}\equiv 0$.

If we perform a straightforward `quantization' of (\ref{sec=tpp}) by replacing the momentum by derivative,
\begin{eqnarray}\label{p--d}
 p_{{\Sigma}} \mapsto -i \partial_{{\Sigma}} \; , \qquad
\end{eqnarray}
consider (\ref{sec=tpp}) as a (first class) constraint and impose
its quantum version as a condition on the wave function, we clearly arrive at the weak version (\ref{sec=tdd=w}) of the section condition.

However, let us imagine that we first solved the classical mechanic relation (\ref{sec=tpp}) and shown that its general solution have the form
\begin{eqnarray}\label{py=Kpr}
 p_{{\Sigma}} = K_{\Sigma}{}^r p_r\; , \qquad r=1,..., \tilde{n}\;
\end{eqnarray}
with some set of independent (or constrained among themselves) $p_r$, $r=1,...,\tilde{n}$.
Then, performing the quantization of (\ref{py=Kpr}) with a prescription of (\ref{p--d}) (and
assuming that $p_r$'s are conjugate to some subset $y^r$ of central charge coordinates $y^\Sigma$) we arrive at the solution (\ref{dy=Kdr}) of the strong section conditions\footnote{Probably the understanding of this fact was also a motivation beyond the recent study in \cite{Cederwall:2015jfa}.}. Moreover, on this way,
one can help to find the general solution of the strong section conditions. Below we will find this for the $E_{7(+7)}$ EFT and show that this general solution is described by the expression of the type
(\ref{py=Kpr}) modulo the $SU(8)$ transformations.

The above discussion motivated us to search for supersymmetric mechanics models in $\Sigma^{(60|32)}$ which might generate `classical section conditions' (\ref{sec=tpp}). Moreover, it is not difficult to conclude that if such a hypothetical model is found in curved $\Sigma^{(60|32)}$, that interesting property should be preserved in flat superspace.

In this paper we present a superparticle model in flat central charge superspace $\Sigma_0^{(60|32)}$ which generates a part of (\ref{sec=tpp}). It can be considered as an improved version of model proposed by de Azc\'arraga and Lukierski in 1986 \cite{deAzcarraga:1984hf} which we also discuss below. The fact that the model produces only a part of  CSCs  (\ref{sec=tpp}), $t_{H }{}^{{\Sigma}{\Pi}} \, p_{{\Sigma}}  \, p_{{\Pi}}=0$ involving the generators of a maximal compact subgroup $H=SU(8)$ of the $E_{7(+7)}$ group,  suggested us that probably the  general solution of this part gives also the general solution of the complete set of the  CSCs  (\ref{sec=tpp}). We show that this is indeed the case so that the independent part of the section conditions of the $E_{7(+7)}$ exceptional field theory are the conditions corresponding to the generators of $H=SU(8)$ subgroup of $E_{7(+7)}$ .  It will be interesting to understand whether this situation is also reproduced in the case of other $E_{n(n)}$ exceptional field theories.

\section{Preliminaries}
\subsection{Section conditions of $E_{7(+7)}$ exceptional field theory}

As we have already said, the exceptional field theory
 which has  a manifest $E_{7(+7)}$ symmetry ($E_{7(+7)}$ EFT) \cite{Hohm:2013pua,Hohm:2013uia,Godazgar:2014nqa} is defined in spacetime with $60=4+56$ bosonic coordinates,  of which $4$ are the usual D=4 spacetime coordinates,
$x^\mu$, and the remaining $56$,  $y^{\Sigma}$, can be considered as a  vector in the  fundamental representation of $E_{7(+7)}$. Already in early 80th  such $y^{\Sigma}$   had been introduced in the superfield approach to maximal D=4 supergravity   \cite{Howe:1980th,Howe:2015hpa} as coordinates of central charge superspace, so that from now on will call them  'central charge coordinates' \footnote{These coordinates   can also be extracted from the set of D=11 tensorial central charge coordinates  \cite{vanHolten:1982mx,Curtright:1987zc,Bandos:1998vz,Bandos:2005rr} which in their turn can be considered as  finite subset of the infinite set of tensorial coordinates which appeared in \cite{West:2003fc} in the frame
 of $E_{11}$ proposal \cite{West:2001as,West:2014eza}.}.

In the EFT framework the dependence of the fields on  central charge coordinates is restricted by the {\it strong section conditions}. In the case of $E_{7(+7)}$ ETF these read \cite{Hohm:2013pua,Hohm:2013uia,Godazgar:2014nqa}
\begin{eqnarray}\label{sec=tdFdF2=0}
& t_{G}{}^{{\Sigma}{\Pi}}\partial_{{\Sigma}}  F_1 \, \; \partial_{{\Pi}} F_2=0 \; , \qquad \\
\label{sec=Sp2=0}
&  \Xi^{{\Sigma}{\Pi}}\partial_{{\Sigma}}  F_1 \, \; \partial_{{\Pi}} F_2=0 \; . \qquad
\end{eqnarray}
Here  $F_1$ and $F_2$ are  arbitrary 'physical'  functions of the coordinates $y^\Sigma$ of extended spacetime,
$\Xi{}^{{\Pi}{\Lambda}}=- \Xi{}^{{\Lambda}{\Pi}}$ is the  invariant tensor of $Sp(56)$ (symplectic `metric'), $\Lambda , \Pi, \Sigma=1,...,56$; $t_{G}{}^{{\Sigma}{\Pi}} =
\Xi{}^{{\Pi}{\Lambda}}t_{G\; {\Lambda}}{}^{{\Sigma}}$  where   $t_{G\; {\Lambda}}{}^{{\Sigma}}$  are $E_{7(+7)}$ generators in {\bf 56} representation, $G=1,..., 133$.
A compact way of writing Eq. (\ref{sec=tdFdF2=0}), (\ref{sec=Sp2=0}) is given above in (\ref{sec=tdd=0}) and  (\ref{sec=Sp=0}). In EFT  every physical function $F$ also obeys  a weak section condition
\begin{eqnarray}\label{sec=tddF=0}
& t_{G}{}^{{\Sigma}{\Pi}}\partial_{{\Sigma}} \partial_{{\Pi}} F=0
& . \qquad
\end{eqnarray}
This is a counterpart of (\ref{sec=tdFdF2=0}) with both derivatives applied to one function  $F$; clearly the weak counterpart of (\ref{sec=Sp2=0}) is satisfied identically.

The fundamental {\bf 56} representation  of  $E_{7(+7)}$ can be decomposed on the sum of ${\bf 28}$ and $\overline{{\bf 28}}$ representations of its
maximal compact subgroup $SU(8)$, so that the natural splitting of the set of additional bosonic coordinates is
\begin{eqnarray}\label{yS=yby}
 y^{\Sigma} =  (y^{ij}, \bar{y}_{ij})\; , \qquad y^{ij}=-y^{ji}\; , \qquad (y^{ij})^*=  \bar{y}_{ij}\; , \qquad i,j=1,...,8 \;  .
\end{eqnarray}
This splitting was implied when we called $y^{\Sigma}$  central charge coordinates.
The corresponding decomposition of derivatives is
 \begin{eqnarray}\label{pyS=pypby}
 \partial_{\Sigma}:= \frac {\partial }{\partial y^{\Sigma} } = (\bar{\partial}_{ij} , {\partial}^{ij} ) \; , \qquad \bar{\partial}_{ij} y^{kl}= \delta_{[i}{}^k \delta_{j]}{}^l = \frac 1 2 \left(\delta_{i}{}^k \delta_{j}{}^l- \delta_{j}{}^k \delta_{i}{}^l\right)\; . \qquad
\end{eqnarray}

The adjoint {\bf 133} representation of  $E_{7(+7)}$ can be decomposed as
{\bf 133}={\bf 70}+{\bf 63}; in particular the set of generators of $G=E_{7(+7)}$  splits onto the set of 63
$H=SU(8)$-generators $t_H= t_{i}{}^j$ and 70 $G/H=E_{7(+7)}/SU(8)$ generators $t_{G/H}$
 \begin{eqnarray}\label{tS=tH+tG-h}
t_{G} &=& (t_{G/H}, t_{H}) = (t_{ijkl}, t_{i}{}^j)\; , \qquad \nonumber \\  && (t_{ijkl})^\dagger = \bar{t}^{ijkl}= 1/4! \epsilon^{ijklpqrs}t_{pqrs}\; , \qquad  (t_{i}{}^j)^\dagger=t_j{}^i\; , \qquad  t_{i}{}^i=0
\; . \qquad
\end{eqnarray}
Using this splitting we can write the strong section conditions  (\ref{sec=tdd=0}) as
\begin{eqnarray}\label{sec=G-H}
\partial_{[ij} \otimes \partial_{kl]} - {1\over 4!}\epsilon_{ijkli'j'k'l'} \bar{\partial}{}^{i'j'}\otimes \bar{\partial}{}^{k'l'} =0 \; , \qquad \\ \label{sec=H} \partial_{ik} \otimes  \bar{\partial}{}^{jk} +
\bar{\partial}{}^{jk}  \otimes  \partial_{ik} -{1\over 4} \delta_i{}^j  \partial_{kl} \otimes  \bar{\partial}{}^{kl}  =0 \; . \qquad
\end{eqnarray}
$E_{7(+7)}$ is the subalgebra of $Sp(56)$ and the above representation of the $E_{7(+7)}$  generator is given for the case of the following representation of the symplectic metric $\Xi^{\Sigma\Pi}=
\left(\begin{matrix} 0 & I_{28\times 28} \cr -I_{28\times 28} & 0\end{matrix}\right)$. Hence the   $Sp$-part of string section condition,
(\ref{sec=Sp2=0}), reads
\begin{eqnarray}\label{sec=Sp2} \partial_{ij} \otimes  \bar{\partial}{}^{ij}-   \bar{\partial}{}^{ij} \otimes  {\partial}{}_{ij}=0 \; . \qquad
\end{eqnarray}

The  strong section conditions (\ref{sec=Sp2})-- (\ref{sec=H})  allow for dependence of all the fields on $7$ additional coordinates; the corresponding solution describes embedding of the standard 11D supergravity \cite{Cremmer:1978km} in the $E_{7(+7)}$ exceptional field theory \cite{Hohm:2013pua,Hohm:2013uia}. In \cite{Hohm:2013pua,Hohm:2013uia} it was also described a solution of section with functions depending on 6 additional bosonic coordinates, which corresponds to embedding of 10D type II supergravity in the exceptional field theory (see also recent \cite{Ciceri:2014wya}). Below we will show that there are no solution of the strong section conditions with functions depending on more than $7$ additional coordinates.

\subsection{Central charge superspace}

We denote the local coordinates of the central charge superspace $\Sigma^{(60|32)}$ by
\begin{eqnarray}
\label{calZ=Z+y}
{\cal Z}^{\cal M}= (X^{\mathfrak{m}}, {\vartheta}^{\underline{\check{\alpha}}})=
(x^m, y^{{\Sigma}}, {\vartheta}^{\underline{\check{\alpha}}})\; , \qquad
\end{eqnarray}
and the supervielbein forms of  $\Sigma^{(60|32)}$ by
\begin{eqnarray}
\label{calE=EA+}
E^{\cal A}&:=& d {\cal Z}^{\cal M} E_{\cal M}^{\cal A} ({\cal Z})=
 (E^{\mathfrak{a}} , {E}^{\underline{{\alpha}}} )
 \; , \qquad   E^{\mathfrak{a}} = (E^a, E^{ij}, \bar{E}_{ij}) \; , \qquad
{E}^{\underline{{\alpha}}}= (E^{\alpha}_i, \bar{E}{}^{\dot{\alpha}i})\; , \qquad \\ \nonumber && a=0,1,2,3\, , \qquad
 {\alpha}=1,2\, , \qquad  \dot{\alpha}=1,2\; , \qquad i,j=1,...,8\, . \qquad
\end{eqnarray}
The $SL(2,{\bb C})\otimes SU(8)$ covariant derivatives of the supervielbeins define the torsion two forms
\begin{eqnarray}
\label{TcA:=} T^{\cal A}= {\cal D}E^{\cal A}&=& (T{}^{a},
 T^{ij},  \bar{T}_{ij};  T{}^{\alpha}_{i},  T{}^{\dot{\alpha} i})=
 \frac 12  E^{\cal C}\wedge E^{\cal B}T_{{\cal B}{\cal C}}{}^{\cal A}\; , \\
\label{Ta:=N8+cc} T{}^{a} & := & {\cal D} E{}^{a} =  d E{}^{a} - E{}^{b}\wedge \Omega_b{}^a \; ,
\\
\label{cTij:=N8}
  T^{ij}
& = & {\cal D} E{}^{ij} = d E{}^{ij} +  2E{}^{k[i}\wedge \Omega_k{}^{j]}\; , \qquad
\\
\label{cbTij:=N8}  \bar{T}_{ij} & := & {\cal D} \bar{E}{}_{ij} =  d\bar{E}{}_{ij} + 2
\Omega_{[i}{}^k  \wedge E_{j]k}\; , \qquad ,
\\
\label{Tfa:=N8+cc} T{}^{\alpha}_{i} & :=& {\cal D} E{}^{\alpha}_{i} =
 d E{}^{\alpha}_{i} -  E{}^{\beta}_{i}\wedge \Omega_{\beta}{}^\alpha  - \Omega_{i}{}^j \wedge E{}^{\alpha}_{j} \; , \qquad
\\
\label{Tbfa:=N8+cc} T{}^{\dot{\alpha} i} & :=& {\cal D} E{}^{\dot{\alpha} i} = d E{}^{\dot{\alpha} i}-   E{}^{\dot{\beta} i} \wedge \Omega_{\dot{\beta}}{}^{\dot{\alpha}}-
 E{}^{\dot{\alpha} j} \wedge \Omega_{j}{}^i\; ,
\end{eqnarray}
where $\Omega_j{}^i = -(\Omega_i{}^j)^*= d {\cal Z}^{\cal M} \Omega_{{\cal M}\, j}{}^i  ({\cal Z})$ is the
$SU(8)$ connection superform ($\Omega_j{}^j=0$),  $\Omega^{ab} = \Omega^{ba}= d {\cal Z}^{\cal M} \Omega_{\cal M}^{ab}  ({\cal Z})$ is the spin connection,  and  $ \Omega_{\beta}{}^\alpha=\frac 14 \, \Omega^{ab} \sigma_{ab\, \beta}{}^\alpha = (\Omega_{\dot{\beta}}{}^{\dot{\alpha}})^*$.

The torsion constraints describing, when supplemented by the conditions (\ref{dy=0}), the  ${\cal N}=8$ $D=4$ supergravity as a model in central charge superspace can be found in (the appendices of) \cite{Howe:1980th} and \cite{Howe:2015hpa} (in this latter ${\cal D}_{ij} (...)=0$ and ${\cal D}^{ij} (...)=0$ were used instead of (\ref{dy=0})).

For (most of) our purposes here it is sufficient to discuss the case of  flat central charge superspace. In the flat limit the connections are trivial and  can be set to zero, $\Omega^{ab}=0$, $\Omega_i{}^j=0$, and the fermionic coordinates carry  indices of $SL(2,{\bb C})\times SU(8)$,
\begin{eqnarray}
\label{undTh=th-bth}
\vartheta^{\underline{\check{\alpha}}}=
({\theta}^{\alpha}_{i}, \bar{\theta}^{\dot{\alpha}i})\; , \qquad \alpha =1,2\, ,\qquad \dot{\alpha}=1,2\; , \quad i=1,...,8\; . \qquad
\end{eqnarray}
In  this case it also makes sense the decomposition (\ref{yS=yby}) of $56$ additional coordinates on to conjugate sets carrying indices of the {\bf 28} and $\overline{\bf 28}$ representation of $SU(8)$ and we can use the following set of supervielbein forms
\begin{eqnarray}\label{cEA=flat} {E}^{a}=dx^a - {i\over 2 }d\theta_i\sigma^a\bar{\theta}{}^i +{i\over 2 }\theta_i\sigma^a d\bar{\theta}{}^i \; , \qquad {E}^\alpha_{i}=
d{\theta}^{\alpha}_{i}  \; , \qquad \bar{{E}}^{\dot{\alpha}i} =
d\bar{{\theta}}{}^{\dot{\alpha}i}  \; , \qquad  \\ \label{cEij=flat} {E}^{ij}= dy^{ij} - i d\bar{\theta}^{\dot{\alpha} [i} \, \bar{\theta}_{\dot{\alpha}}^{j]} \; , \qquad  \bar{{E}}_{ij}= d\bar{y}{}_{ij}- i
d{\theta}^{\alpha}_{ [i} \, {\theta}_{j]\alpha}  \; . \qquad
 \end{eqnarray}
These  are invariant under rigid ${\cal N}=8$ $D=4$ supersymmetry generated by supercharges
\begin{eqnarray}\label{Qali=}
Q_{\alpha}^i =- i \partial_{\alpha}^i -\frac 12 (\sigma^a\bar{\theta}^i)_\alpha \partial_a
- {\theta}_{\alpha j} \partial^{ij}\; , \qquad
\bar{Q}_{\dot{\alpha}i} =- i \bar{\partial}_{\dot{\alpha}i} -\frac 12 ({\theta}_i\sigma^a)_{\dot{\alpha}}  \partial_a
- \bar{\theta}^j_{\dot{\alpha}} \bar{\partial}_{ij}
\;  \qquad
 \end{eqnarray}
the superalgebra of which has the form (\ref{SUSYal}) with $P_a=i\partial_a$,
$Z^{ij}=2i\partial^{ij}$ and $\bar{Z}_{ij} =2i\bar{\partial}_{ij}$.

The supervielbeine   (\ref{cEA=flat})
can be also called Cartan forms as  they obey  the torsion `constraints' with constant coefficients
\begin{eqnarray}
\label{Ta0=N8} T{}^{a} & = &
 -iE^{\alpha}_{i}\wedge \bar{E}{}^{\dot{\beta}i}\sigma^{a}_{\alpha\dot{\beta}}, \; , \qquad
\\
\label{cbTij0=N8}  \bar{T}_{ij} & = &  -i E^{\alpha}_{i} \wedge E^{\beta}_{j}\epsilon_{\alpha\beta}\; ,
\qquad
  T^{ij}= -i E^{\dot{\alpha}i} \wedge E^{\dot{\beta}j}\epsilon_{\dot{\alpha}\dot{\beta}} \; ,
\\
\label{Tfa0=N8} T{}^{\alpha}_{i} & :=& 0\; , \qquad T{}^{\dot{\alpha} i} = 0\; , \qquad
\end{eqnarray}
which can be identified with structure equations of a supergroup manifold associated to the supersymmetry algebra (\ref{SUSYal}), {\it i.e.} of the flat ${\cal N}=8$ central charge superspace $\Sigma_0^{(60|32)}$.

\section{A superparticle model in  ${\cal N}=8$ central charge superspace}

An interesting superparticle model in flat ${\cal N}=8$ superspace enlarged by central charge  coordinates,  $\Sigma_0^{(60|32)}$,
was proposed by de Azc\'arraga and Lukierski in  sec.  2 of
\cite{deAzcarraga:1984hf}. In this section, after reviewing this model we will present its
improved version, which does possess the $\kappa$--symmetry without imposing any condition by hand.
In the next sec. 4, using suggestions provided by this superparticle model we will analyze the interdependence of the section conditions and present their  general solution.

\subsection{De Azc\'arraga--Lukierski model}

In \cite{deAzcarraga:1984hf} Jos\'e de Azc\'arraga and Jerzy Lukeirski discussed (among others) the following action for massive superparticle in  ${\cal N}=8$ $D=4$ central charge  superspace $\Sigma^{(60|32)}$:
\begin{eqnarray}\label{S=AL86}
S= m\int d\tau \sqrt{\hat{E}^a_\tau \hat{E}_{a\tau} }  + 2b \int d\tau \sqrt{\hat{E}^{ij}_\tau \hat{\bar{E}}_{ij\, \tau} }\; .
\end{eqnarray}
In it $m$ is a constant with dimension of mass, $b$ is a constant obeying
\begin{eqnarray}\label{b2=}
b^2= {\cal N} m^2/4\; =\; 2m^2
\end{eqnarray}
(the reason to impose this relation will be clear from the discussion below)
and
\begin{eqnarray}\label{hEa=}
\hat{E}^{{\cal A}} = d\tau \hat{E}_\tau^a = d\hat{{\cal Z}}{}^{{\cal M}}(\tau) {E}_{{\cal M}}{}^{{\cal A}}(\hat{{\cal Z}}(\tau)) \;
\end{eqnarray}
is the pull back of the supervielbein form (\ref{calE=EA+}) ((\ref{cEA=flat}) in the case of flat superspace $\Sigma_0^{(60|32)}$) to the superparticle worldline $W^1$ which is described as a line in $\Sigma^{(60|32)}$ by the coordinate functions $\hat{{\cal Z}}{}^{{\cal M}}(\tau)$.

It is convenient to introduce the canonical momenta
\begin{eqnarray}\label{pga=papij}
p_{\mathfrak{a}} &:=&  \frac {{\cal L}} {\partial \hat{E}^{\mathfrak{a}}_\tau}=  (p_a , \bar{p}_{ij} , p^{ij}) \; , \qquad  \\
\label{pa=cEa}
& & p_a = m {\hat{E}_{a\tau} \over \sqrt{\hat{E}^a_\tau \hat{E}_{a\tau} }}
\; , \qquad
 \bar{p}_{ij} = b{
 \hat{\bar{E}}_{ij\, \tau} \over \sqrt{\hat{E}^{ij}_\tau \hat{\bar{E}}_{ij\, \tau} }}\; , \qquad
  p^{ij}= b{
\hat{E}^{ij}_\tau \over \sqrt{\hat{E}^{ij}_\tau \hat{\bar{E}}_{ij\, \tau} }}\; . \qquad
\end{eqnarray}
It is not difficult to check that  these obey the constraints
\begin{eqnarray}\label{p2=m2}
p_ap^a=m^2\; \qquad \\
\label{pbp=b2}
\bar{p}_{ij} {p}^{ij} =b^2= {\cal N} m^2/4\; =\;  2m^2\; . \qquad
\end{eqnarray}

Then the fermionic variation of the action (\ref{S=AL86}) can be written in the form
\begin{eqnarray}\label{vS=AL86}
\delta_\epsilon S= \int d\tau  p_{\mathfrak{a}} \hat{E}^{{\cal B}}_\tau \epsilon^{\underline{\alpha}} T_{\underline{\alpha}{\cal B}}{}^ {\mathfrak{a}} (\hat{{\cal Z}})\; , \qquad
 \end{eqnarray}
where $\epsilon^{\underline{\alpha}}= (\epsilon_i^{{\alpha}}\; , \bar{\epsilon}{}^{\dot{\alpha}i})$ are the parameters of the fermionic variations defined by
\begin{eqnarray}\label{fvEal}
\epsilon^{\underline{\alpha}}=(\epsilon_i^{{\alpha}}\; , \bar{\epsilon}{}^{\dot{\alpha}i})= i_\epsilon \hat{E}{}^{{\underline{\alpha}}}:=
\delta_\epsilon\hat{{\cal Z}}{}^{{\cal M}}(\tau) {E}_{{\cal M}}{}^{{\underline{\alpha}}}(\hat{{\cal Z}}(\tau)) \; , \qquad \\ \label{fvEa}
 i_\epsilon \hat{E}^{{\mathfrak{a}}}:=
\delta_\epsilon\hat{{\cal Z}}{}^{{\cal M}}(\tau) {E}_{{\cal M}}{}^{^{{\mathfrak{a}}}}(\hat{{\cal Z}}(\tau))=0 \; . \;
 \end{eqnarray}

In the case of flat central charge superspace $\Sigma^{(60|32)}_0$, the supervielbein forms are given in
(\ref{cEA=flat})
and
the variation of the action  (\ref{vS=AL86}) is
\begin{eqnarray}\label{vS=AL86-0}
\Sigma^{(60|32)}_0\; : \qquad \delta_\epsilon S= -2i \int d\tau  \left(p_a(\sigma^a \hat{\bar{E}}^{i}_\tau)_\alpha  + 2 p^{ij}\hat{E}_{\alpha j\, \tau}\right) \delta \theta_i^{{\alpha}} + c.c. \; .
 \end{eqnarray}
Now one can easily check that \cite{deAzcarraga:1984hf}
\begin{eqnarray}\label{ep=kap}
\delta_{\kappa} \theta^{\alpha}_i= p_a\bar{\kappa}_{\dot{\alpha}i}\tilde{\sigma}^{a\dot{\alpha}\alpha} - 2\bar{p}_{ij} \kappa^{\alpha j}=  (\bar{\epsilon}{}^{\dot{\alpha}i})^*\; , \qquad \delta_{\kappa} \bar{\theta}{}^{\dot{\alpha}i}= p_a\tilde{\sigma}^{a\dot{\alpha}\alpha} \kappa_{\alpha}^i - 2{p}^{ij} \bar{\kappa}_{\dot{\alpha}j}
 \end{eqnarray}
would be a gauge symmetry of the action {\it provided} the momenta in central charge directions were restricted by the condition
\begin{eqnarray}\label{pbp=Ipp}
\bar{p}_{ik}p^{kj}= - \delta_i{}^j m^2/4 \; .  \qquad
 \end{eqnarray}
This implies $\bar{p}_{ik}p^{ik}= {\cal N}/2  p_ap^a= 2 m^2$, so that one sees that the relation (\ref{b2=}) between parameters of the action (\ref{S=AL86}) is also related to the wish to have the $\kappa$--symmetry.

De Azc\'arraga and Lukierski appreciated themselves that imposing the condition (\ref{pbp=Ipp}) by hand at this stage does not look  consistent (although in the present perspective one can find it similar to imposing the section condition in EFT)   and, after making this interesting observation,  passed in \cite{deAzcarraga:1984hf} to the study of different, more conventional  models. In the next sec. \ref{improvedS} we will present an improved superparticle model which does not need imposing constraint by hand to possess $\kappa$--symmetry with (\ref{ep=kap}).

\subsection{Improved superparticle model in central charge superspace}
\label{improvedS}

The above de Azc\'arraga--Lukierski model inspired us to study the following first order action
\begin{eqnarray}\label{S=impr}
S=\int d\tau (p_a \hat{E}_\tau^a + \bar{p}_{ij}\hat{E}_\tau^{ij}+p^{ij} \hat{\bar{E}}_{ij\, \tau})   - \int d\tau \Lambda_j{}^i \left(\bar{p}_{ik} {p}^{kj}+{m^2\over 4}\delta_i{}^j\right) - \int d\tau  {1\over 2}e (p_ap^a- m^2) \; . \nonumber \\ {}\end{eqnarray}
Here the momenta $p_a$, $p^{ij}$ and $ \bar{p}_{ij}$ are independent variables,
$\hat{E}_\tau^a$,  $\hat{E}_\tau^{ij}$ and $\hat{\bar{E}}_{ij\, \tau}$ are the pull--backs ((\ref{hEa=})) of the flat supervilebien forms  (\ref{cEA=flat}) and (\ref{cEij=flat}); finally  $\Lambda_i{}^j $ and $e$ are
Lagrange multipliers introduced to impose the constraints  (\ref{pbp=Ipp}) and (\ref{p2=m2}).

The action is invariant under the $\kappa$--symmetry (\ref{ep=kap}) supplemented by the Lagrange multiplier transformations:
 \begin{eqnarray}\label{vTh=kap}
\delta\theta^{\alpha}_i= p_a\bar{\kappa}_{\dot{\alpha}i}\tilde{\sigma}^{a\dot{\alpha}\alpha} - 2\bar{p}_{ij} \kappa^{\alpha j}\; , \qquad \delta\bar{\theta}{}^{\dot{\alpha}i}= p_a\tilde{\sigma}^{a\dot{\alpha}\alpha} \kappa_{\alpha}^i - 2{p}^{ij} \bar{\kappa}_{\dot{\alpha}j}\; , \qquad \\ \label{vL=kap}
 \delta \Lambda_i{}^j =4i(\partial_\tau \theta^{\alpha}_i\, \kappa_{\alpha}^j+ \partial_\tau \bar{\theta}{}^{\dot{\alpha}j}\bar{\kappa}_{\dot{\alpha}i})\; , \qquad
\delta e =-2i(\partial_\tau \theta^{\alpha}_i\, \kappa_{\alpha}^i+ \partial_\tau \bar{\theta}{}^{\dot{\alpha}i}\bar{\kappa}_{\dot{\alpha}i})\; .
 \end{eqnarray}

Notice that this model is not apparently  equivalent to the one described by the second order action (\ref{S=AL86}). Indeed, the equations of motion for the internal momenta produce the relations
\begin{eqnarray}\label{cEA=LpA}
\hat{E}^{ij}_\tau= \Lambda_k{}^{[i}p^{j]k}\; , \qquad \hat{\bar{E}}_{ij\, \tau} = \Lambda_{[i}{}^k\bar{p}_{j]k} \;  \qquad
\end{eqnarray}
 which differ from (\ref{pa=cEa})
(while $p_a$ equations gives the standard $\hat{E}^{a}_\tau= ep_a$). The Lagrange multiplier variations produce the constraints (\ref{p2=m2}) and (\ref{pbp=Ipp}) and, using these one can  deduce from
(\ref{cEA=LpA}) the following interesting interrelation between pull--backs of the Cartan forms
\begin{eqnarray}\label{cE=ppcbE}
\bar{p}_{ik}\bar{p}_{jl}\hat{E}^{kl}=  \frac {m^2}4 \hat{\bar{E}}_{ij} \; , \qquad {p}^{ik}{p}^{jl} \hat{\bar{E}}_{kl}  =   \frac {m^2}4 \hat{E}^{ij}\; , \qquad
{p}^{ik} \hat{\bar{E}}_{kj}=  -\hat{E}^{ik}\bar{p}_{kj}\; .
\end{eqnarray}

However, despite the lack of apparent equivalence, the fermionic variation of  (\ref{S=impr}) is still described by (\ref{vS=AL86}) and, in the case of generic curved target superspace,   by (\ref{vS=AL86-0}).

\section{Superparticle and section conditions}

Notice that Eq. (\ref{pbp=Ipp}) provides us with a counterpart of  $SU(8)$ part of the set of section conditions (\ref{sec=H}). To be precise, if section conditions appeared from the superparticle model,
in the classical approximation the derivatives would be replaced by particle momenta so that we would have the constraint (\ref{sec=tpp}), the  SU(8) part of which  coincides with  (\ref{pbp=Ipp}),
\begin{eqnarray}\label{SU=sec}
\bar{p}_{ik}p^{kj}+  {1\over 8}\, \delta_i{}^j \bar{p}_{kl}p^{kl}=0 \; , \qquad
 \end{eqnarray}
while the remaining $E_{7(+7)}/SU(8)$ part reads
\begin{eqnarray}\label{E77-SU=Sec}
\bar{p}_{[ij}  \bar{p}_{kl]} - {1\over 4!}\epsilon_{ijkli'j'k'l'} p^{i'j'}p^{k'l'} =0 \; . \qquad
\end{eqnarray}
The classical mechanics  counterpart of the section condition  involving the Sp(56) metric,  (\ref{sec=Sp2}),  is obeyed identically.

As we have already discussed in the Introduction, the straightforward quantization of the constraints (\ref{SU=sec}) and (\ref{E77-SU=Sec}) by a simple prescription $( \bar{p}_{ij}, {p}^{ij}) \mapsto (\bar{\partial}_{ij}, {\partial}^{ij})$ results in the the weak section conditions (\ref{sec=tdd=w}) only. However,   if we first find the general solution of  (\ref{SU=sec}) and (\ref{E77-SU=Sec}) expressing  $( \bar{p}_{ij}, {p}^{ij})$ in terms of smaller set of momenta $p_r$, and then quantize this, supplementing
$( \bar{p}_{ij}, {p}^{ij}) \mapsto (\bar{\partial}_{ij}, {\partial}^{ij})$ by $p_r\mapsto {\partial}_r$, we arrive at the general solution of the strong section conditions (\ref{sec=tdFdF2=0}).

The superparticle model discussed in the previous section generates the SU(8) part of the classical section conditions only. Then the  association of  the classical section conditions (\ref{SU=sec}) and (\ref{E77-SU=Sec}) with symmetry generators of $H=SU(8)$ and  $G/H=E_{7(+7)}/SU(8)$ makes tempting to search the explanation of the  appearance of (\ref{SU=sec}) constraint only on the basis of the fact that
 $E_{7(+7)}$  symmetry of  ${\cal N}=8$ supergravity \cite{Cremmer:1979up}  is realized  dynamically in such a way that in its flat superspace limit only the symmetry under the $SU(8)$ subgroup of $E_{7(7)}$ `survives'.
Namely, the superspace formulation of ${\cal N}=8$ supergravity \cite{Brink:1979nt,Howe:1981gz} involves an $E_{7(+7)}$-valued matrix `bridge' superfield
\begin{eqnarray}
\label{V=inE7}
 {\cal V}^{{\Sigma}}{}_{({\Pi})} = ({\cal V}^{{\Sigma}}{}_{kl}, \bar{{\cal V}}{}^{{\Sigma}\, kl } )
\quad \in \quad E_{7(+7)}\;  \qquad
\end{eqnarray}
which is transformed by  right multiplication on a matrix in fundamental representation of $E_{7(+7)}$ and by left multiplication on a matrix  in a reducible {\bf 56}={\bf 28}+$\overline{\bf 28}$ representation of $SU(8)$. In flat superspace limit one usually sets $ {\cal V}^{{\Sigma}}{}_{({\Pi})} = ({\delta}^{{\Sigma}}{}_{kl}, \delta{}^{{\Sigma}\, kl } )$ which is preserved by $SU(8)$ subgroup of the above $E_{7(+7)}\otimes SU(8)$ transformations. This implies that in flat superspace limit the (rigid nonlinearly realized) $E_{7(+7)}$ symmetry of ${\cal N}=8$ supergravity reduces to  $SU(8)$.\footnote{To be more  precise, one can state that
$E_{7(+7)}$ symmetry is  realized trivially in flat superspace limit. Indeed, in this limit the matrices
 $ {\cal V}^{{\Sigma}}{}_{({\Pi})}$ can take any constant value in $E_{7(+7)}$ group; the  $E_{7(+7)}$ symmetry is realized by multiplication on these matrices. However, as the derivative of constant matrices vanish and only these, through the $E_{7(+7)}$ Cartan forms, influence the superspace geometry \cite{Brink:1979nt},    the flat superspace  coordinates, as well as any dynamical system described in terms of these,    do  not see any effect of the  $E_{7(+7)}/SU(8)$ transformations.
 }

Then  the superparticle model in flat superspace is manifestly  invariant under nontrivial transformations of $SU(8)$ symmetry, and, from this point of view it looks natural  that it generates as  constraints only a part of section condition involving the $SU(8)$ generator. However, accepting this argument (which is not a proof, but rather a speculation), one cannot conclude  that a curved superspace generalization of the superparticle model, which is expected to see the $E_{7(+7)}$ symmetry, may generate additional {\it nontrivial} constraints corresponding to the coset $G/H=  E_{7(+7)}/SU(8)$. Indeed, if this were the case, it would  contradict the physical requirement of non-singularity of the flat superspace limit.

The above discussion suggests  that the appearance of only a $SU(8)$ part (\ref{SU=sec})   of the section conditions (\ref{sec=tpp}) may reflect the fact that their  G/H part (\ref{E77-SU=Sec}) is dependent, i.e. is satisfied as a consequence of its $H=SU(8)$ part (\ref{SU=sec}).
In other words, it inspires to check whether the general solution of all the set of section conditions
coincide with the general solution of the $SU(8)$ part of the section conditions. We are going to show that this is indeed the case.

Let us begin by discussing the solution of Eq. (\ref{SU=sec}) (or equivalently, Eq. (\ref{pbp=Ipp})).
First notice that  it has a solution with 7 real parameters $p^I=(p^I)^*$:
\begin{eqnarray}\label{pij=pIgg8}
\bar{p}_{ij}= p^I (\gamma^I\tilde{\gamma}^8)_{ij}= p^{ij}\; , \qquad I=1,...,7\; .
 \end{eqnarray}
In it $\gamma^{\check{I}}_{p\dot{q}}=\tilde{\gamma}{}^{\check{I}}_{\dot{q}p}= (\gamma^{{I}}_{p\dot{q}}, \gamma^{8}_{p\dot{q}})$  are the $SO(8)$ Clebsch-Gordan coefficients.  If we wont to solve
 (\ref{pbp=Ipp}), the seven real $p^I$  should obey \begin{eqnarray}\label{pI2=m2}
 p^Ip^I=m^2/4\; , \qquad
 \end{eqnarray}
but this is not essential for the conclusion on interdependence of a part of section conditions.

Actually, the solution (\ref{pij=pIgg8}) can be written in terms of  SO(7) gamma matrices $\Gamma^I_{pq}$
 \begin{eqnarray}\label{pij=pIGI}
\bar{p}_{ij}= p^I \Gamma^I_{ij}= p^{ij}\; , \qquad I=1,...,7\; ,
 \end{eqnarray}
the representation of which in terms of $SO(8)$  Clebsch-Gordan coefficients,
$\Gamma^I_{pq}= (\gamma^I\tilde{\gamma}^8)_{pq}$,
 has been used above.

At this stage one can wonder whether (\ref{pij=pIgg8}) solves also the G/H part of the classical section conditions, Eq. (\ref{E77-SU=Sec}).
This is not manifest from the first glance as for this to be the case the following identity for
SO(8) Klebsh-Gordan coefficients should hold
\begin{eqnarray}\label{SO8gId}(\gamma^{(I|}\tilde{\gamma}^8)_{[pq} (\gamma^{|J)}\tilde{\gamma}^8)_{rs]}\; = {1\over 4!} \epsilon_{pqrsp'q'r's'}(\gamma^{(I|}\tilde{\gamma}^8)_{[p'q'} (\gamma^{|J)}\tilde{\gamma}^8)_{r's']}\; . \qquad
 \end{eqnarray}
As the condition  for SO(7) gamma matrices $\Gamma^I_{pq}$ it reads
\begin{eqnarray}\label{SO7gId}\Gamma^{(I|}_{[pq} \Gamma^{|J)}_{rs]}\; = {1\over 4!} \epsilon_{pqrsp'q'r's'}\Gamma^{(I|}_{[p'q'} \Gamma^{|J)}_{r's']}\; . \qquad
 \end{eqnarray}

However, a more careful study of the solution (\ref{pij=pIgg8})  shows that it is equivalent to the classical counterpart of the solution to the whole set of section conditions found by Hohm and Samtleben in
\cite{Hohm:2013pua,Hohm:2013uia}. This was written `in $SL(7)$ basis' and reads \begin{eqnarray}
\label{H+S=sol}
q^{pq}= 2\delta^{[p}_I\delta^{q]}_8 p^I\; , \qquad  \tilde{q}_{pq}=0
\;  , \qquad \end{eqnarray}

In this paper we, following \cite{Cremmer:1979up}, work in the  `$SU(8)$ basis'. The relation between components of  $SL(8,{\bb R})$ and $SU(8)$ decompositions of the fundamental representations of  $E_{7(+7)}$ can be described by  (see e.g. \cite{Kallosh:2012yy})    \begin{eqnarray}
\label{SU-SL}
\left(\begin{matrix} p^{ij} {} \cr \bar{p}_{ij} \end{matrix} \right)= {\cal S} \left( \begin{matrix} q^{pq}\cr \tilde{q}_{pq} \end{matrix} \right)
\;  , \qquad
\label{cS=}
{\cal S}={1\over 4\sqrt{2}} \gamma^{ij}_{pq}\otimes \left(\begin{matrix} 1 & i  {} \cr 1 & -i \end{matrix} \right)\; . \qquad
 \end{eqnarray}
Using this we find that  the solution of Hohm and Samtleben  (\ref{H+S=sol}) of the whole set of section conditions of $E_{7(+7)}$ EFT
coincides with (\ref{pij=pIgg8})  when written in the  $SU(8)$ basis. Then, as  Hohm-Samtleben solution  solves
the whole set of section conditions, and (\ref{pij=pIgg8}) is proved to be equivalent to it,
this solves also Eq.
(\ref{E77-SU=Sec}) and thus provides a solution of the complete set of section conditions, (\ref{SU=sec})  and (\ref{E77-SU=Sec}).

Notice that, by pass, we have proved the identities (\ref{SO7gId}) for $d=7$ gamma matrices and  (\ref{SO8gId}) for the $SO(8)$ Clebsch-Gordan coefficients.

Now we are going to  show that (\ref{pij=pIgg8}) describes  general solution of (\ref{SU=sec}) up to $SU(8)$ transformations.
To this end, we first notice that  two solutions of this equation, $p^{ij}_{(1)}$ and  $p^{ij}_{(2)}$, are related by an $U(8)$ transformations. Indeed, as  (\ref{SU=sec}) implies that the matrices  $p^{ij}_{(1)}$ and  $p^{ij}_{(2)}$ are not degenerate, so is $U_i{}^j= \frac {4}{m} \bar{p}_{(1)ik} p^{jk}_{(2)}$ which obeys $UU^\dagger=I$ and hence belongs to $U(8)$. Below we will concentrate on  the $SU(8)$ part of this solution-generating $U(8)$ symmetry of (\ref{SU=sec}), as
only $SU(8)$ part of $U(8)$ transformations of the momenta lives invariant also the constraint
(\ref{E77-SU=Sec}).

Interestingly, just the freedom in  $SU(8)$ transformations is sufficient to describe the general solution of (\ref{SU=sec})  starting from  (\ref{pij=pIgg8}). To prove  this,
let us show that an arbitrary nondegenerate 56-vector $p_\Sigma=(\bar{p}_{ij},   {p}^{ij})$ can be obtained by $SU(8)$ transformation applied to one of the vectors described by the  complexification of
 (\ref{pij=pIgg8}),
 \begin{eqnarray}
\label{p14=sol}
  {p}^{ij}=  w_I \Gamma^I_{ij}\;  , \qquad \bar{p}_{ij}= \bar w_I \Gamma^I_{ij}\; , \qquad
   \end{eqnarray}
with complex $w^I$.


To this end, let us firstly notice  that the  expressions in (\ref{p14=sol}) contain $14$ real parameters (in 7 complex $w^I$)
and are invariant under $SO(7)$ subgroup of $SU(8)$. This latter fact implies that the nontrivial orbit  of (\ref{p14=sol}) under $SU(8)$ is generated by 42 elelments of  the coset  $SU(8)/SO(7)$.
Finally, as $42+14=56$, we conclude that arbitrary
complex $\bar{p}_{ij}$ can be generated by $SU(8)$ transformations starting from  $\bar{p}_{ij}= \bar w_I \Gamma^I_{ij}$ with some complex $w^I$.
Now, $\bar{p}_{ij}= \bar w_I \Gamma^I_{ij}$ solves the constraint (\ref{SU=sec}) only in the case when
$w_I$ is real, $w_I =\bar w_I =p^I$.
As the constraint (\ref{SU=sec}) is $SU(8)$ invariant, we conclude that, modulo $SU(8)$ transformations, its general solution is given by (\ref{pij=pIgg8}).

Furthermore, as (\ref{pij=pIgg8}) solves also (\ref{E77-SU=Sec}) and this latter is $SU(8)$ invariant, we conclude that the $SU(8)$ orbit of  (\ref{pij=pIgg8}) provides us with the general solution of the complete set of classical section conditions (\ref{sec=tpp}) of the $E_{7(+7)}$ EFT.

This  result implies that, modulo $SU(8)$ transformations  (which constitute the gauge symmetry of ${\cal N}=8$ $D=4$ supergravity \cite{Brink:1979nt}), the 'quantized' version of Eq. (\ref{pbp=Ipp}),
\begin{eqnarray}\label{bd=gIdI}
\bar{\partial}_{ij}= (\gamma^I\tilde{\gamma}^8)_{ij}  \partial_I  = \partial^{ij}\; , \qquad
 \end{eqnarray}
provides the general solution for the  complete set of  section conditions (\ref{sec=G-H}) and (\ref{sec=H}).  Of course, it looks desirable to make the above statement on `modulo $SU(8)$ transformations' more concrete  in the quantized version. We hope to turn to this issue in the future.

To conclude we have shown that  in the $E_{7(+7)}$ effective field theory it is not necessary to impose the complete set of section conditions as all the job on restricting the dependence of fields on additional coordinates is done by the part of the section conditions which involves the generator of
$SU(8)$ subgroup of $E_{7(+7)}$.

It will be interesting to understand whether the sets of the section conditions for other $E_{n(+n)}$ exceptional field theories are also reducible.

\section{Conclusion and outlook}

In this paper we have made a stage towards better understanding of the structure of the mysterious section conditions of the exceptional field theories  (EFTs) \cite{Hohm:2013pua,Hohm:2013vpa,Hohm:2013uia,Hohm:2014fxa,Godazgar:2014nqa,Musaev:2014lna,Ciceri:2014wya,Abzalov:2015ega}. Namely, we have shown that in   $E_{7(+7)}$ EFT   it is not necessary to impose the complete set of section conditions as all the job on restricting the dependence of fields on additional coordinates is done by the part of the section conditions which corresponds to $SU(8)$ subgroup of $E_{7(+7)}$. On the way we have also proved that, as might be expected, the general solution of the strong section conditions implies the dependence of the EFT fields on not more than 7 of 56 additional bosonic coordinate.

This progress is reached using the suggestions coming from the superparticle model in central charge superspace. We have discussed how the strong section condition can be reproduced starting from the classical constraints of a superparticle model. In essence, the 'first quantize than solve' way of treating the classical mechanic counterparts of the section conditions reproduce the weak section conditions only, however there exists an alternative 'first solve then quantize' way which allows to reproduce the general solution of the strong section conditions starting from its classical counterparts.

The superparticle model in central charge superspace we have used in our study is an improved version of the de Azc\'arraga--Lukierski model proposed in sec. 2 of \cite{deAzcarraga:1984hf}. This is a massive superparticle model in $D=4$ ${\cal N}=8$ superspace enlarged by 56 central charge coordinates. On first glance, the fact that superparticle is massive (in 4-dimensional perspective, $p_ap^a=m^2\not=0$) might look strange as to obtain supergravity one would rather quantize a massless $D=4$ ${\cal N}=8$ superparticle model \cite{Billo:1999ip}, while the quantization of massive superparticle with $32$ supersymmetries shall inevitably result in appearance of higher spin fields in the spectrum. However,  a more careful thinking suggests to  relate this to the fact that U-duality, which is expected to be manifest in EFT, is the property of M-theory and higher spin fields should appear here as M-theoretic counterpart  of the massive modes of String theory.
Furthermore, one can observe that the conditions for external and internal momenta, which follows from our superparticle model (and which should be partially imposed by hand in the original model of  \cite{deAzcarraga:1984hf}) related the square of external momenta ($p_ap^a=m^2\not=0$) and of the internal momenta ($\bar{p}_{ik}p^{kj}= - \delta_i{}^j m^2/4 $,
see (\ref{pbp=Ipp})) to the same parameter $m^2$. Hence we can treat the superparticle as being massless in an enlarges spacetime. In particular, on the 7-parametric solution (\ref{pij=pIgg8}) of  the classical section conditions  the model should become equivalent to  a massless 11 dimensional superparticle.

One of the natural applications of our approach is in search for superfield formulation of exceptional field theories,   particularly for the  formulation of the
$E_{7(+7)}$ EFT in  curved ${\cal N}=8$ $D=4$  central charge superspace $\Sigma^{(60|32)}$. The basis
of such a formulation can be provided by an appropriate set of the constraints on the generalized torsion (\ref{TcA:=})--(\ref{Tbfa:=N8+cc}).
The requirement of the invariance of the curved superspace generalization of our superparticle model, (\ref{S=impr}) with generic supervielbein  (\ref{calE=EA+}), under  $\kappa$--symmetry, similar to (\ref{ep=kap}) can allow us to find a set of appropriate constraints  (see
\cite{Witten:1985nt,Grisaru:1985fv,Bergshoeff:1985su,Bergshoeff:1987cm} for such approach to D=10 and D=11 supergravity in standard superspaces). An especially interesting question is  whether these still hypothetical constraints, if found,  would describe the $E_{7(+7)}$ EFT when  supplemented by the section conditions or they would produce these as their selfconsistency conditions.
These issues are under investigation.

Another interesting direction for future study is to formulate superparticle models in central charge superspaces appropriate for construction of superfield formulations of the other $E_{n(+n)}$ EFTs, with $n\leq 6$ and with $n=8$. In particular, this might help to understand  whether  the sets of the section conditions for other $E_{n(+n)}$ EFTs are also reducible.

\subsection*{Acknowledgements}
The author is grateful to Tom\'as Ort\'{\i}n for useful discussions and collaboration on early stages of this project, to Henning Samtleben for useful conversation, to Paul Howe for useful communications and to Dima Sorokin for useful discussions. This work has been supported in part by the Spanish MINECO grant FPA2012-35043-C02-01  partially financed  with FEDER funds,  by the Basque Government research group grant ITT559-10 and the Basque Country University program UFI 11/55.

\end{document}